\documentstyle[preprint,aps]{revtex}
\draft

\begin{document}

\title{Influence of growth direction and strain conditions
on the band line-up at GaSb/InSb and InAs/InSb interfaces}
\author{S.Picozzi and A. Continenza}
\address{INFM - Istituto Nazionale di Fisica della Materia\\
Dipartimento di Fisica \\
Universit\`a degli Studi di L'Aquila, 67010 Coppito (L'Aquila), Italy \\ and}
\author{A.J.Freeman}
\address{Department of Physics and Astronomy and Material Research Center\\
Northwestern University, Evanston, IL 60208 (U.S.A.)\\}

\maketitle

\narrowtext

\begin{abstract}
First-principles full potential linearized augmented plane wave
(FLAPW) calculations have been performed for lattice-mismatched
common-atom III-V interfaces. In particular,
we have examined the effects of epitaxial strain and
ordering direction on the valence band
offset in [001] and [111]
 GaSb/InSb and InAs/InSb superlattices, and found that the
valence band maximum is always {\em higher} at the InSb side of the
heterojunction, except for the common-anion system grown on an InSb
substrate.
The comparison between equivalent structures having the same substrate
lattice constant, but different growth axis, shows that
for comparable strain conditions,
the ordering direction slightly influences the band
line-up, due to small differences of the charge readjustment
at the [001] and [111] 
interfaces.
On the other hand, strain is shown to strongly affect
the VBO; 
in particular, as the pseudomorphic growth conditions 
are varied, the bulk contribution to the band line-up changes markedly, whereas
the interface term is almost constant. On the whole,
our calculations  yield a
band line-up  that decreases linearly as the
substrate lattice constant is increased, showing its high tunability
as a function of different pseudomorphic growth conditions.
Finally, the  band line-up at the lattice matched InAs/GaSb interface
determined using the transitivity rule gave
perfect agreement between  predicted and experimental results.
\end{abstract}
\bigskip

\newpage

\section{Introduction}
In the last few years ``band offset engineering" ({\em i.e.}
the possibility of tuning the electronic and transport properties
of semiconductor heterojunctions through modifications of their
valence band offset (VBO) induced
by strain, growth process, number of layers etc.)
has attracted great attention 
for both scientific and technological
reasons \cite{goss}.
Only recently,however, have lattice mismatched heterostructures begun to realize
their potential, due to new developments in preparation techniques
which finally allowed pseudomorphic crystal growth without misfit
dislocations \cite{qwapl}.
In a parallel way, some theoretical works
\cite{dandrea,ale90,VDW,priester,oloumi,taguchi,az94,maria}
focused on strained heterojunctions, but still much effort is needed to
understand what mainly affects the band line-up at the interface.

In this work, we consider homopolar isovalent heterostructures.
In particular,
we examine III-V superlattices (SLs), namely the common-anion system
GaSb/InSb (lattice mismatch of 5.7 $\%$) and the common-cation system
InAs/InSb (lattice mismatch of 6.4 $\%$), concentrating on the effect of
ordering direction and
strain conditions
determined by SL pseudomorphic growth on a given substrate.
{\em Ab-initio} self-consistent full potential linearized augmented
plane wave FLAPW \cite{FLAPW} local density
calculations were performed for
(GaSb)$_3$/(InSb)$_3$  and (InAs)$_3$/(InSb)$_3$  SLs, grown
along the [001]  (tetragonal symmetry) and [111] (trigonal symmetry)
directions, in which
the interface was represented by three  alternating layers of each
binary constituent (12 atoms in each unit cell). In what follows, it
is important to keep in mind that if strain is not taken into account,
the valence bands in unstrained GaSb and InSb are
predicted  to line-up (therefore giving zero VBO),
whereas the unstrained InSb topmost valence band is
expected to be 0.51 eV {\em higher} in energy than that in
unstrained InAs (see Ref.\cite{mailhot} and references therein).

As pointed out by Mailhot and  Smith \cite{ms1,ms2}, [111]
ordered strained-layer SLs show large polarization fields oriented along the
growth direction, 
which lead to a shift in the electronic energy levels. However,
this effect is negligible for ultrathin SLs, such as
those examined in the present work. Moreover, the effect completely
vanishes for [001] ordered SLs, due to the symmetry properties
of the strain tensor. Finally,
we neglect interdiffusion processes which lead to interfacial composition
changes (i.e. we consider an atomically abrupt geometry) and
relaxations at the interface of the anion-cation
distance (bulk bond length away from the interface
are considered equal to those immediately next to it):
this, in fact, is expected \cite{ale90,duke,foulon}
to introduce little modification
of the charge rearrangement at the junction and hence of the VBO.

\section{Computational and structural parameters}
Most of the computational parameters are common to those used
previously for [001]
and [111] ordered (1x1)
 SLs \cite{silvia}, except for the wave function cut-off
($k_{max}$ = 2.7 a.u.).[17]
Tests performed by increasing the $k_{max}$ up to 3.1 showed
a change in the VBO of less than 0.01 eV.
There is, however, an important difference with respect to the calculations
performed for ultrathin ($n$ =1)
SLs regarding the cation $d$-shell:
in the present work, the Ga 3$d$ and In 4$d$ electrons are
treated as part of the core and not as valence electrons (previous works
\cite{ale90} demonstrated a slight dependence (about 0.03 eV) of the VBO
in common-cation systems
on the $d$-shell treatment) - which results in a strong reduction of the
computational effort. Furthermore, the core charge spilling out of the Ga
and In
muffin tin spheres was treated using an exact overlapping charge method, thus
minimizing the error introduced by the treatment of semicore states.

The structural parameters (reported elsewhere \cite{silvia}) are determined
according to the macroscopic theory of elasticity (MTE), whose validity in
predicting the correct structure for the determination of the VBO is
established \cite{maria}. Our choice is also justified by the results obtained
in the case of ultrathin SLs \cite{silvia}, which
were found to be in good agreement with
those obtained from total energy minimization.

In order to study the dependence of the VBO on the strain,
we examined different strain conditions for (AC)$_3$/(BC)$_3$-type
[001] ordered SLs (the common cation case
(AB)$_3$/(AC)$_3$ is treated in an analogous way):
(i) pseudomorphic growth of a BC epilayer on an AC substrate; (ii) ``free
standing mode", equivalent to a system grown on an A$_{0.5}$B$_{0.5}$C
substrate (denoted
in the following as ``Av. subs."); 
and (iii) pseudomorphic growth of an AC epilayer on a BC substrate.

On the other hand, the dependence of the VBO on the ordering direction is
studied through a comparison of the [001] and [111] ordered SLs
grown on a fixed substrate with average lattice constant,
 but different crystallographic
orientations. This choice of the substrate 
and the consequent lattice relaxation, leads to
 a small difference (about 1.4 $\%$) between the
 lattice constants of the binary constituents along the [001] and  [111]
 growth
direction.

\section{Results and discussion}
In analogy with the common experimental approach followed in photoemission
measurements, we have evaluated the VBO using core electron binding
energies
as reference levels \cite{mmf}. We have chosen the
$s$-levels of the common atom C
({\em i.e.} Sb in the common-anion system and In in the common-cation
system). Note
 that other choices of core levels for different atoms ({\em
i.e.} Ga and In in the common-anion system, As and Sb in the common-cation
system) would produce a VBO value differing
 from those reported here by at most 0.06  eV, which has thus to be
considered as our numerical uncertainty.

The calculation of the VBO, $\Delta\:E_v$, is done according to the
following expression:
\begin{equation}
\Delta\:E_v\:=\:\Delta\:b+\Delta\:E_b
\label{equazione}
\end{equation}
where the interface term $\Delta\:b$ indicates the relative
core level alignment
of the two C atoms at opposite sides of the interface
(one belonging to the AC side and the other to the BC side), while
$\Delta\:E_b$ indicates the binding energy difference (relative to the
valence band maximum (VBM)) of the same core levels evaluated in the  binary
constituents, opportunely strained to reproduce the
elastic conditions of the SL.

First of all, we focus our attention on the elastically relaxed
``Av. subs."
[001] and [111] ordered SLs.
We should now notice that there are two inequivalent interfaces
along the [111] growth axis \cite{resta}; for example, in the common-anion SL,
we can have the ordering direction parallel either to the InSb interface bond
or to the GaSb interface bond. Our results indicate that the $\Delta\:b$ term
is essentially the same (within 0.02 eV) in the two different
situations, suggesting that the effect due to the particular
geometry at the interface is very small.

Table \ref{vbo1} lists the
contributions due to the interface ($\Delta\:b$) and to the
strained bulks ($\Delta\:E_b$) and the
resulting values of the VBOs ($\Delta\:E_v$)
as a function of the ordering direction; the
superscripts $(nr)$ and $(r)$ indicate respectively the non-relativistic
and relativistic ({\em i.e.} spin-orbit coupling
treated in a perturbative approach)
calculations.

We note that the interface term has a positive sign, indicating that the
Sb-core levels are deeper at the GaSb (InAs) side of the
common-anion (common-cation) interface, compared to the corresponding
levels at the InSb side. Further, the first contribution ($\Delta\:b)$
is seen to be
sensitive to the crystallographic ordering (the two values for
[001] and [111] growth axis differ  by 0.1 eV both in the common-anion
and the common-cation systems), while
the second contribution to the VBO ($\Delta\:E_b$)
is almost uninfluenced by the ordering direction.
On the whole, we do not observe a marked dependence of the VBO
on the crystallographic
ordering at the interface.
Since it is well-known for lattice matched structures
 \cite{nato,vdw1} that the band line-up is independent of
interface orientation, the VBO change we find in going
from the [001] to the [111] ordered SLs has to be related
only to the appreciable mismatch which causes a different relaxation
of the interface bond-lengths.

Let us now look at the role of the mismatch in determining the
band line-up. Our results for [001] systems with different
pseudomorphic growth conditions
are shown in Table \ref{vbo2}, where the notation is analogous
to that of Table \ref{vbo1}.
Note that the interface term $\Delta\:b$ is very similar in the same
SL grown on the three different substrates, implying that
the charge readjustment at the interface is almost independent of
the strain conditions.
This is confirmed by the results obtained with a small change (by as much
as 0.6 $\%$) of the bond length at the interface: the calculated $\Delta\:b$
is consistent with the one obtained for MTE structures within 0.03 eV.

On the other hand,
the $\Delta\:E_b$ term ({\em i.e.} the bulk contribution to the VBO)
varies dramatically, showing that
 the core level binding energies in the strained
 binary suffers an appreciable change
when growing the SL on different
substrates.
In fact, the energy of the topmost valence level (and hence the binding
energy $E_b$) is determined by  the interplay of the
spin-orbit coupling and  the non-cubic ``crystal field" \cite{azapl}.
In particular, the second
of these two effects is critically dependent on strain conditions,
and is thus the  origin of the large
difference between the $\Delta \:E_b$ in Table \ref{vbo2}.
Furthermore, what is remarkable about Table \ref{vbo2}
is the clear trend shown by
the VBO as a function of the substrate lattice constant: the smaller
the $a_{sub}$, the more  the InSb topmost
valence level is raised with respect to the VBM of the other SL constituent.

In order to understand more fully the action of the
strain on the VBO, we have also estimated the band line-up with respect
to unstrained binaries for the [001] interfaces. Thus, we
substituted in Eq.(\ref{equazione}) the $\Delta\:b$ value obtained for [001]
interfaces (which was found to be almost independent
of strain effects) and the $\Delta\:E_b$, evaluated
starting from the zincblende bulk unstrained constituents
(i.e. disregarding the effect of strain on the binary's VBM).
Taking into account the spin-orbit coupling, we obtain
$\Delta\:E_v^{rel}$ = 0.03 eV and
$\Delta\:E_v^{rel}$ = 0.49 eV for the GaSb/InSb and InAs/InSb
heterojunctions, respectively. These results match perfectly
 those reported in  Ref. \cite{mailhot} and
the above mentioned 
valence bands perfect
alignment for the common anion system and
$\Delta\:E_v^{rel}$ = 0.51 eV for the
common cation case,
showing that, if strain is not correctly
taken into account, completely different results are found.

It is important to notice that the strain acting on the energy
of the topmost valence band level is also responsible for the spatial
localization of this state. We find, in fact, that in
all  of the common anion (common cation) structures,
the hole carriers are mainly localized on the InSb side of the
heterojunction, while in the GaSb/InSb system grown on an InSb substrate
we find a complementary situation.
This is clear from the decreasing trend of the VBO as the lattice
constant is increased and from the sign change in the InSb substrate case
(showing that the VMB in GaSb is higher in energy than in InSb).

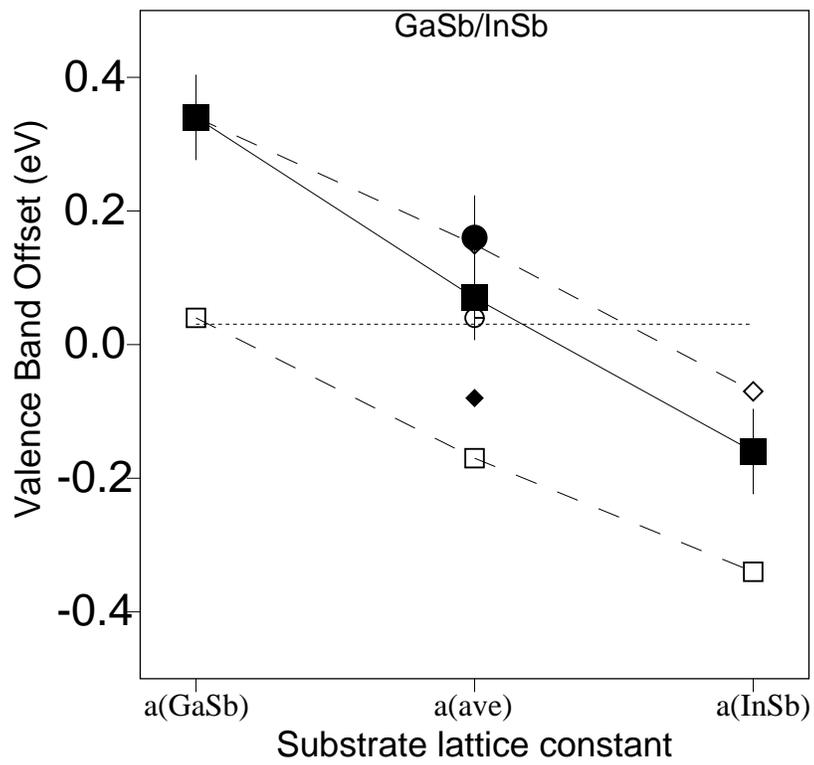
\begin{figure}
\caption{Valence band offset (in eV) for GaSb-InSb SLs as a function of the
substrate lattice parameter. Our results (together with their error bars)
are evidenced by filled squares ([001] SLs)
or filled circles ([111] SLs) and solid line. 
The dotted line shows the behaviour of unstrained GaSb and InSb.
Empty squares:
Ref. [23]; empty diamonds: Ref. [6]; filled diamond:
Ref. [24]; empty circles: Ref. [10].}
\label{fig1}
\end{figure}

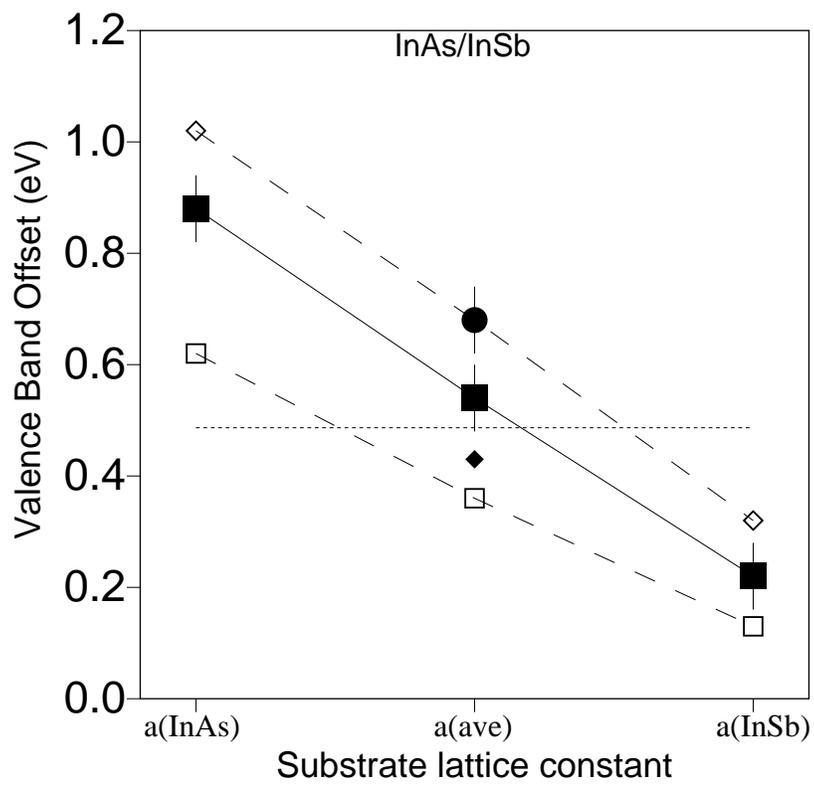
\begin{figure}
\caption{Valence band offset (in eV) for InAs/InSb SLs as a function of the
substrate lattice parameter. 
Symbols are the same as those in Fig. 1.}
\label{fig2}
\end{figure}

Figures \ref{fig1}  and  \ref{fig2},   for common-anion and
common-cation  interfaces
respectively, illustrate the  linear dependence
(see the solid line in the figures)
of the VBO on the lattice parameter which determines the SL pseudomorphic
growth. Thus, the two figures show that the GaSb/InSb and InAs/InSb SLs
provide a
good opportunity for tuning their VBO: a range of about
0.5 eV for common-anion
and of 0.7 eV for common-cation systems is covered by varying
the strain conditions determined by the substrate.

Let us now compare our results with other theoretical predictions,
obtained from model \cite{VDW,cardona},
semi-empirical \cite{ichii} and {\em ab-initio} \cite{az94} calculations, as
illustrated in Figures \ref{fig1} and \ref{fig2}.
Note that all the predicted values agree  with those of the present
work (except those of Ref. \cite{cardona}), within their uncertainty of a few hundredths of an eV \cite{VDW}
and our error bars, respectively. 
Incidentally, we observe that a similar  disagreement between  {\em ab initio}
  results
and those obtained by Cardona and Christensen \cite{cardona} was also
 found 
in other III-V isovalent heterojunctions, such as GaP/InP
\cite{az94} and GaAs/InAs \cite{maria}. 
Furthermore, the linear trend of the band offset as a function of the strain
 found in the present work   is in
excellent agreement with the  predictions of other theoretical work
\cite{VDW,cardona} and is reasonably expected to reproduce the real
situation.

So far, we have completely omitted a discussion of the conduction band
offset ($\Delta\:E_c$), due to well known failures of LDA in
predicting the correct band gap energies. However, we should now
point out that, using an  approximate estimate of empirical band gaps
(see Ref. \cite{silvia} for details), we can obtain information
on the different kinds of band line-ups
as growth conditions are changed.
In fact, we find a type I alignment for all the [001] common anion
interfaces, while [111] GaSb/InSb shows a type II staggered alignment
 (partial
overlap of the band gaps).
On the other hand, for [001] common cation grown both on an InAs and on an
average substrate, we find a type II broken gap line-up, with the InAs
conduction band minimum lower in energy than the InSb VBM. Finally, 
the [001] InAs/InSb grown on InSb and the [111] InAs/InSb heterojunctions
contain a semimetallic compound (InAs), thus leading to
a type III alignment.

Starting from the results obtained from {\em ab initio}
calculations
\cite{az94,maria,ale90,bibbia} for different
[001] oriented strained layer interfaces (with
similar lattice mismatch and grown on a  substrate having lattice
constant averaged over those of the constituents),
it is interesting to discuss the VBO
trend as a function of the atomic species involved.
We compare the value obtained for the common anion GaSb/InSb
(lattice mismatch of 5.7 $\%$ and $\Delta\:E_v$ = 0.07 eV, in good
agreement with $\Delta\:E_v$ = 0.04 eV reported in
 Ref. \cite{az94}), with those obtained
for GaP/InP \cite{az94} (lattice mismatch of 7.4 $\%$ and 
 $\Delta\:E_v$ =
0.01 eV) and  for GaAs/InAs \cite{maria}
(lattice mismatch of 5.7 $\%$ and $\Delta\:E_v$ = 0.00 eV).
 Taking into account  that these results are
 obtained by different computational methods and are therefore
 affected by different error bars, it appears 
 that, under similar strain conditions,
 the VBO is almost uninfluenced
by the change of both anions
in the common anion systems.
Therefore, since the $\Delta\:b$ term is expected to be constant (due
to similar ionicity difference of the constituents), these results
seem to suggest that also the $\Delta\:E_b$ term has to be similar in all
the GaX/InX (X=P, As, Sb) structures.
A similar $\Delta\:E_v$ is also found, if we compare our results for InAs/InSb
(lattice mismatch of 6.4 
$\%$ and $\Delta\:E_v$ = 0.54 eV) with
 GaAs/GaSb (lattice mismatch of 7.2 $\%$), where both cations
are changed ($\Delta\:E_v$ = 0.65 $\pm$ 0.1 eV \cite{bibbia}).
This observation can be explained, considering the
 anionic character (see Ref. \cite{az94},
 Ref. \cite{silvia} and references therein)
of the topmost valence level in the binary constituents, which determines
the binding energy contribution ($\Delta\:E_b$) to
the valence band offset:  for example,
provided that the state of strain is similar in the constituent materials
grown on an average substrate,
we don't expect
a strong difference in the band line-up
if we change the cation at both sides of the  common cation interface.

From the experimental point of view, the high mismatch (about 6 $\%$)
between the lattice constants of the two SL constituents results
in great difficulty to grow the SL systems,
without misfits and dislocations.
Only recently,  an InSb quantum-well has been realized in GaSb
\cite{qwapl};
starting from photoluminescence peak emission energy data and from
calculations based on a standard finite square-well model
\cite{bastard,alavi} (taking
into account strain \cite{asai,ji}), a VBO of 0.16 eV was obtained
that is
quite different from the one, 0.34 eV, reported in Table \ref{vbo2}.

Unfortunately,
 a similar disagreement 
between theoretical and experimental data, obtained using different
 techniques,  was found also for other
homopolar isovalent III-V interfaces, such as the GaAs/InAs (see Ref.
\cite{maria}
and references therein)
and GaP/InP ($\Delta\:E_v$ = 0.01 eV, the theoretical result \cite{az94}
against
$\Delta\:E_v$ = 0.60 eV, the experimental results from photoluminescence
\cite{armelles})  systems,  that are reasonably close to the common-anion
one studied here.
In particular, in a recent work 
focused on GaAs/InAs SLs, Ohler {\em et al.} 
\cite{ohler} obtained (from ultraviolet
photoelectron spectroscopy measurements of the cation $d$-core levels) a
$\Delta\:E_v$ value in disagreement (by as much as 0.3 eV) with theoretical
predictions \cite{VDW,priester,taguchi,maria}.
Notwithstanding some differences between the experimental and theoretical
 VBO values, many important
observations  discussed above are confirmed by Ohler {\em et al.}
experimental work
\cite{ohler}. 
For
example, the linear trend found \cite{ohler} for $\Delta E_v$
(in GaAs/InAs SLs) as a function of
$a_{sub}$ agrees with  theoretical predictions \cite{maria}
and with our results; further, the
independence of the $\Delta_b$ term on the strain conditions
discussed above,
is experimentally confirmed  \cite{ohler} by the trend of the In 4$d_{5/2}$
and Ga 3$d_{3/2}$ core-level binding energy
difference, which is 
almost unaffected by the different substrate used in growing the
heterostructure.

As a last comment, we think that it  worthwhile to remark
 that the linear trend found in Ref. \cite{maria} 
for the GaAs/InAs VBO's as a function of the growth conditions leads
to a slope that is almost equal to that of Fig. \ref{fig1} for GaSb/InSb
systems.

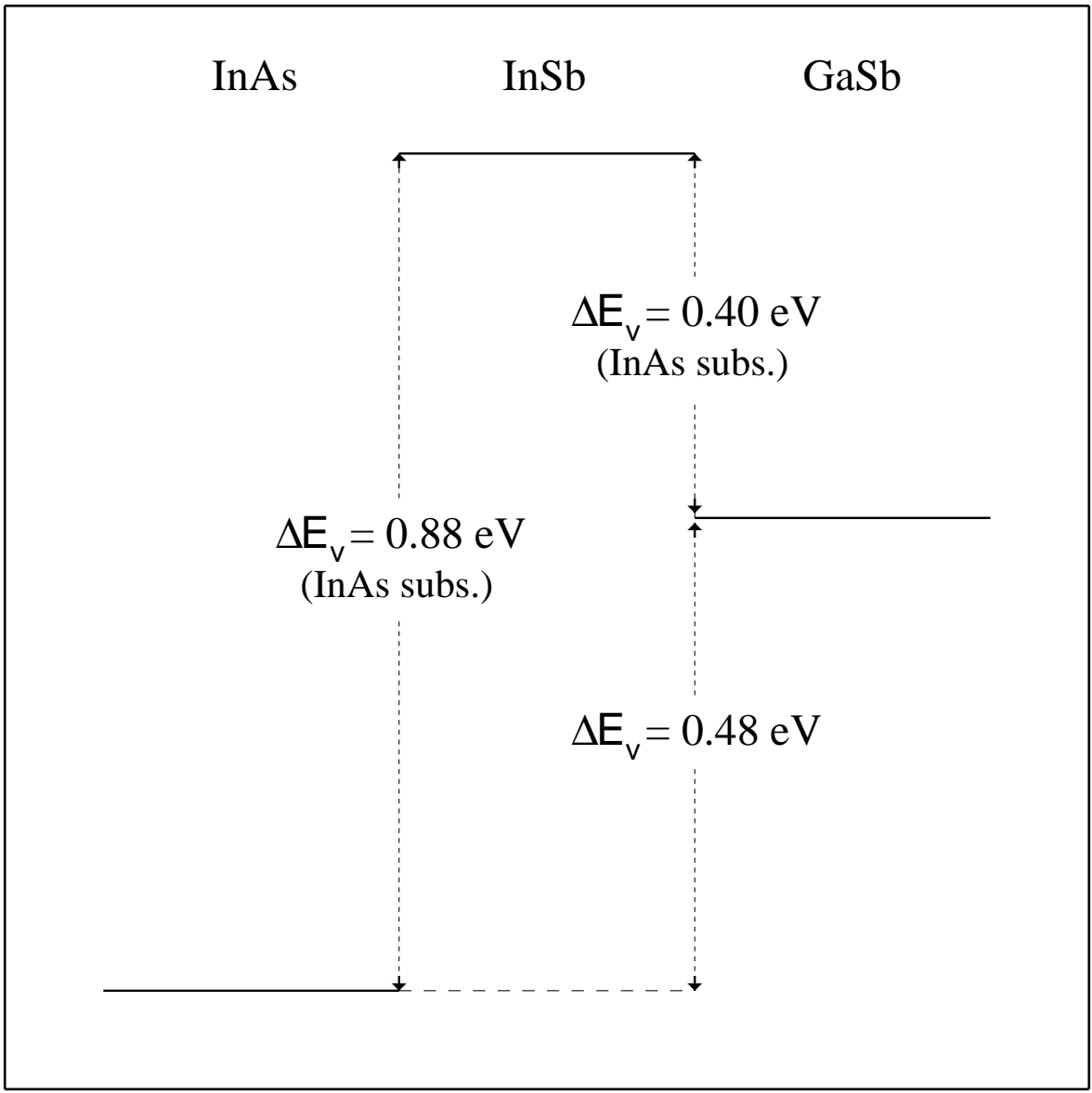
\begin{figure}
\caption{Valence band offset (in eV) for InAs/GaSb
interface, obtained using the transitivity rule.}
\label{fig3}
\end{figure}

Starting from the band offset transitivity rule - which is well established
(to within 0.02 eV) \cite{foulon}
for [001] common-atom superlattices - we  can derive the VBO for the InAs/GaSb
system, an almost lattice matched interface (the two lattice
constants differ only by 0.6 $\%$). This system is attracting
more and more attention recently for its unusual type II broken-gap band
line-up.
Through a linear interpolation of the common-anion band offsets
as a function of the substrate lattice constant, we have calculated
the VBO for the GaSb/InSb system as if grown on an InAs substrate.
As evidenced above (see Fig. \ref{fig1}),
the linear approximation is expected to be reasonably valid;
furthermore, in this case the extrapolation is obtained
for a lattice constant ($a_{InAs}$) which differs by only 0.7 $\%$ from
one of our
self-consistent results ($a_{GaSb}$).
We thus report in Fig. \ref{fig3} our calculated
VBO for the common-cation SL on an InAs substrate and
the extrapolated value for the ideal
common-anion SL grown on an InAs substrate. Using the transitivity rule,
we obtain:

\begin{center}
$\Delta\:E_v (InAs/GaSb) = \Delta\:E_v (InAs/InSb)_{InAs-subs.}
- \Delta\:E_v(InSb/GaSb)_{InAs-subs.}$
\end{center}
so that $\Delta\: E_v (InAs/GaSb)_{InAs-subs.}$ = 0.88 - 0.40 = 0.48 eV.
This result is in good agreement with the available experimental values
(0.46 eV \cite{jaros}, 0.51 eV \cite{kroemer}).

\section{Conclusions}
In summary, we have studied the valence band offsets in [001] and
[111] GaSb/InSb
and InAs/InSb interfaces by means of {\em ab-initio} FLAPW calculations,
focusing our attention on its dependence on
ordering direction and strain conditions.
Our results indicate
that, under the same strain conditions,
 the former has quite a small effect on the band line-up
mainly due to the different  
structural relaxation of the interface atoms at the [001]
and [111] heterojunctions.
On the other hand, a much more important effect is due to
pseudomorphic growth on different substrates:
the high tunability of the VBO (about 0.5 eV and 0.7 eV
for common-anion  and for common-cation SLs, respectively)
is evidenced
by its linear decreasing trend as the substrate lattice
constant is increased, mainly due to the bulk contribution
to the band line-up.
Finally, the transitivity rule was used to determine the InAs/GaSb
valence band offset and good agreement between theory and experiment
was obtained.

\section{ACKNOWLEDGEMENTS}
We thank B. W. Wessels and M. Peressi 
for stimulating discussions and a careful
reading of the manuscript. Useful discussions with S. Massidda are
also acknowledged. Work at Northwestern University supported
by the MRL Program of the National Science Foundation, at the Materials
Research Center of Northwestern University, under Award No.
DMR-9120521,
and by a grant of computer time at the NSF
supported Pittsburgh Supercomputing Center.
Partial support by a supercomputing grant at Cineca (Bologna, Italy)
through the Consiglio Nazionale delle Ricerche (CNR) is also acknowledged.

\begin{table}
\centering
\caption{Interface term ($\Delta\:b$), strained bulk term
($\Delta\:E_b$) and
valence band offset ($\Delta\:E_v$)  for elastically relaxed
(GaSb)$_3$/(InSb)$_3$
and (InAs)$_3$/(InSb)$_3$ -average substrate- 
superlattices as a function of the
ordering direction ($\Delta\:E_b^{(nr)}$ and
$\Delta\:E_v^{(nr)}$) and including ($\Delta\:E_b^{(r)}$ and
$\Delta\:E_v^{(r)}$) spin-orbit
effects). Energy differences (in eV)
 are considered positive if the level relative
to the InSb layer is higher in energy with respect to
 the GaSb (InAs) layer
in the common-anion (common-cation) system.}
\vspace{5mm}
\begin{tabular}{|cl|r|cc|cc|} 
\centering
& & $\Delta\:b$ & $\Delta\:E_b^{(nr)}$ & $\Delta\:E_b^{(r)}$ &
$\Delta\:E_v^{(nr)}$ & $\Delta\:E_v^{(r)}$ \\ \hline
(GaSb)$_3$/(InSb)$_3$ &[001] & +0.19 & -0.21 & -0.12 & -0.02 & +0.07 \\
&       [111] & +0.29 & -0.27 & -0.13 & +0.02 & +0.16 \\ \hline
(InAs)$_3$/(InSb)$_3$ &[001] & +0.07 & +0.27 & +0.47 & +0.34 & +0.54\\
& [111] & +0.17 & +0.30 & +0.51 &+0.47 & +0.68\\
\end{tabular}
\label{vbo1}
\end{table}

\begin{table}
\centering
\caption{Interface term ($\Delta\:b$), strained bulk term
($\Delta\:E_b$) and
valence band offset ($\Delta\:E_v$)  for (GaSb)$_3$/(InSb)$_3$
and (InAs)$_3$/(InSb)$_3$ [001] superlattices as a function of the
substrate lattice parameter neglecting
($\Delta\:E_b^{(nr)}$ and $\Delta\:E_v^{(nr)}$) and including
($\Delta\:E_b^{(r)}$ and $\Delta\:E_v^{(r)}$) spin-orbit
effects). Energy differences (in eV)
 are considered positive if the level relative
to the InSb layer is higher in energy with respect to
 the GaSb (InAs) layer
in the common-anion (common-cation) system.}
\vspace{5mm}
\begin{tabular}{|cl|r|cc|cc|} 
\centering
& & $\Delta\:b$ & $\Delta\:E_b^{(nr)}$ & $\Delta\:E_b^{(r)}$ &
$\Delta\:E_v^{(nr)}$ & $\Delta\:E_v^{(r)}$ \\ \hline
&GaSb-subs.& +0.20 & +0.10 &+0.14&  +0.30 & +0.34\\
(GaSb)$_3$/(InSb)$_3$ &Av. subs. & +0.19 & -0.21 & -0.12 & -0.02 & +0.07 \\
&InSb-subs. & +0.18 & -0.47 & -0.34 & -0.29 & -0.16\\ \hline
&InAs-subs. & +0.09 & +0.65 & +0.79 & +0.74 & +0.88\\
(InAs)$_3$/(InSb)$_3$ &Av. subs. & +0.07 & +0.27 & +0.47 & +0.34 & +0.54\\
& InSb-subs. & +0.05 & -0.04 & +0.18 & +0.01 & +0.22\\
\end{tabular}
\label{vbo2}
\end{table}

\end{document}